# Quasi-linear magnetoresistance and paramagnetic singularity in Hypervalent Bismuthide

Zhongchen Xu[1,2,3], Yi Yan[3,4], Zhihao Liu[3,4], Jie Pang[2,3], Guohao Dong[2,3], Xiutong Deng[2,3], Shengnan Zhang[5], Xianmin Zhang[1], Youguo Shi[2,3,6], and Quansheng Wu[3,4,6]

[1]*Key Laboratory for Anisotropy and Texture of Materials (Ministry of Education), School of Material Science and Engineering, Northeastern University, Shenyang 110819, China*

[2]*Center of Materials Science and Optoelectronics Engineering, University of Chinese Academy of Sciences, Beijing 100190, China*

[3]*Beijing National Laboratory for Condensed Matter Physics and Institute of Physics, Chinese Academy of Sciences, Beijing 100190, China*

[4]*School of Physical Sciences, University of Chinese Academy of Sciences, Beijing 100190, China*

[5]*Beijing Polytechnic College, Beijing, 100042, China*

[6]*Songshan Lake Materials Laboratory, Dongguan, Guangdong 523808, China*

Correspondence: Xianmin Zhang (zhangxm@atm.neu.edu.cn), Youguo Shi (ygshi@iphy.ac.cn) and Quansheng Wu (quansheng.wu@iphy.ac.cn)

Zhongchen Xu and Yi Yan contributed equally to this work

## Abstract

Materials featuring hypervalent bismuth motifs have generated immense interest due to their extraordinary electronic structure and exotic quantum transport. In this study, we synthesized high-quality single crystals of $La_3ScBi_5$ characterized by one-dimensional hypervalent bismuth chains and performed a systematic investigation of the magnetoresistive behavior and quantum oscillations. The metallic $La_3ScBi_5$ exhibits a low-temperature plateau of electrical resistivity and quasi-linear positive magnetoresistance, with anisotropic magnetoresistive behaviors suggesting the presence of anisotropic Fermi surfaces. This distinctive transport phenomenon is perfectly elucidated by first-principles calculations utilizing the semiclassical Boltzmann transport theory. Furthermore, the nonlinear Hall resistivity pointed towards a multiband electronic structure, characterized by the coexistence of electron and hole carriers, which is further supported by our first-principles calculations. Angle-dependent de Haas-van Alphen oscillations are crucial for further elucidating its Fermiology and topological characteristics. Intriguingly, magnetization measurements unveiled a notable paramagnetic singularity at low fields, which might suggest the nontrivial nature of the surface states. Our findings underscore the interplay between transport phenomena and the unique electronic structure of hypervalent bismuthide $La_3ScBi_5$, opening avenues for exploring novel electronic applications.

# INTRODUCTION

Magnetoresistance (MR), defined as the variation in resistance under an applied magnetic field, serves as a crucial probe into Fermi surface properties and holds promise for applications in magnetic sensors and hard disk drives.[1, 2] In the realm of extreme quantum limits, where the magnetic field quantizes orbital motion, the dominance of the lowest Landau level (LL) typically leads to the expectation of pronounced quasi-linear MR at high fields.[3, 4] The exotic MR behavior in Dirac and Weyl materials has led to a flurry of extensive research, including significant negative linear MR in Dirac semimetals $Cd_3As_2$[5] and $Na_3Bi$,[4] and titanic MR in Type II Weyl semimetal $WTe_2$.[6] These Dirac and Weyl fermions in those topological materials also manifest various exotic quantum phenomena, including chiral anomaly and nontrivial quantum oscillations[7, 8]. In addition, several non-magnetic materials with zero-gap properties show large unsaturated positive MR, the possible mechanism is based on the classical theory of long transport mean free path caused by mobility distribution or the quantum effect with linear dispersive band structure.[9, 10] Consequently, MR will exhibit unconventional behaviors across diverse novel materials[11, 12], thereby fostering the development of potential electronic devices.

The "heavy" element bismuth and its compounds are predestined for topological properties induced by spin-orbit coupling (SOC).[13] The $AMnBi_2$ ($A$ = Sr, Ca, Eu, or Yb) systems[14, 15] exhibit fascinating properties that derive from their building unit, a two-dimensional Bi square net containing relativistic fermions. In parallel, the $Ln_3MX_5$ system ($Ln$ = rare earth; $M$ = transition metals; $X$ = As, Sb, Bi) is predicted to be a topological candidate owing to its linear Bi chains. The quasi-one-dimensional chains in $Sm_3ZrBi_5$ give rise to dispersive linear and flat bands, while Bi-Bi interaction results in topological band crossing.[16, 17] Experimental verification of nontrivial topological bands in the $Ln_3MX_5$ family has been successfully demonstrated via de Haas van Alphen (dHvA) oscillation analysis in $La_3MgBi_5$.[18, 19] These compounds are intriguing due to their narrow-gap topological bands influenced by SOC, potentially exhibiting novel electronic transport properties when subjected to a magnetic field. Notably, research on the MR of the low-dimensional $Ln_3MX_5$ system remains limited.

In this work, we report details of single-crystal growth of hypervalent bismuthide $La_3ScBi_5$ with one-dimensional Bi chains from the $Ln_3MX_5$ family. Density functional theory calculations indicate that SOC significantly affects $La_3ScBi_5$, as the band structure near the Fermi energy is predominantly comprised of Bi $p$ orbitals. The electrical transport and magnetic properties were comprehensively investigated, leading to a study of the electron band structure in both the bulk and surface regions. The bulk crystal shows that the magnetoresistance exhibits a crossover between a conventional quadratic law and an unusual linear dependence, along with highly anisotropic electronic transport properties. Analysis of the Hall effect reveals that the Fermi surface of $La_3ScBi_5$ consists of multiple pockets arising from a multiband electronic structure. Even $La_3ScBi_5$, which has quasi-one-dimensional features and pronounced anisotropy, has a distinctly three-dimensional electronic structure, as evidenced by its angle-dependent dHvA oscillations. Notably, the presence of a prominent paramagnetic singularity suggests the potential nontrivial nature of its surface states. Our study employs La$_3$ScBi$_5$ as a model system to investigate hypervalent bismuthides, focusing on the quasi-linear MR and paramagnetic singularities observed in electron-rich motifs.

## RESULTS

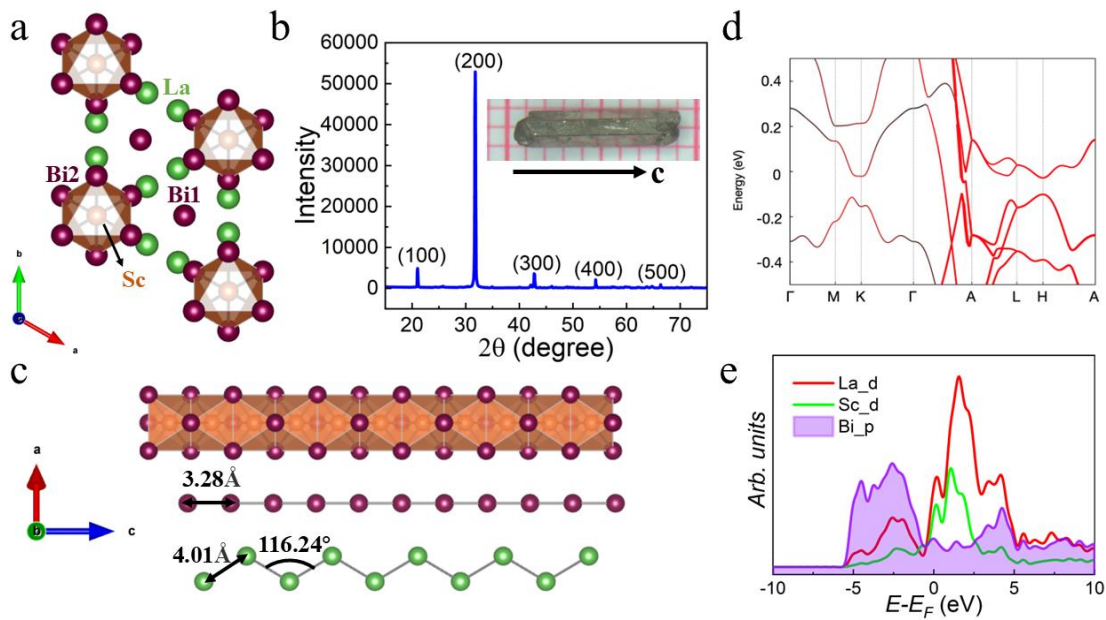

**Fig. 1 Crystal structure of La$_3$ScBi$_5$. a** Schematic diagram of crystal structure with a view along the c-axis projection. **b** Single-face X-ray diffraction pattern for a flux-grown single crystal. The inset presents an optical photograph of a typical single crystal. **c** Quasi-one-dimensional patterns in La$_3$ScBi$_5$: The separate one-dimensional (1D) ScBi$_6$ face-sharing octahedral chains, 1D Bi chains, and La zig-zag chains. **d** Electronic band structure of La$_3$ScBi$_5$ with spin-orbit coupling (SOC), which emphasizes the contribution of band projection of Bi chain *p* orbitals. **e** Partial density of states calculations with SOC, highlighting the contribution of Bi *p* states at the Fermi level.

**Quasi-One-Dimensional Patterns of Crystal Structure.**

La$_3$ScBi$_5$ crystallizes in a rod-like, hexagonal structure, characteristic of other members in the *Ln*$_3$*MX*$_5$ family, with a space group of *P*6$_3$/*mcm* (No. 193). Analysis of the single-crystal X-ray diffraction (XRD) data reveals that La$_3$ScBi$_5$ exhibits lattice parameters of ***a*** = 9.746(3) Å and ***c*** = 6.561(3) Å, matching closely with the previous single-crystalline study.[20] The single-crystal XRD data collection and refinement parameters are gathered in Tables SI-III (Supporting Information). The (*h*00) reflections in Fig. 1b demonstrate the excellent crystal quality.

Figure 1c showcases quasi-one-dimensional patterns within La$_3$ScBi$_5$, composed of the separate one-dimensional (1D) ScBi$_6$ face-sharing octahedral chains, 1D Bi chains, and La zig-zag chains aligned along the c axis, rather than forming a perfect 1D system. A notable feature is the Bi-Bi bond length within the chain, measuring approximately 3.28 Å, significantly longer than expected for a localized and covalent Bi-Bi single bond. This elongated yet symmetric bonding can be comprehended by "hypervalent" bonding, wherein a −2 charge on Bi is stabilized by the delocalization of electrons within the chain.[21] The presence of extended hypervalent bonds necessitates an electron count of one electron per bond, resulting in half-filled bonds crucial for stabilizing symmetry-protected band crossings.[22-24] Moreover, hypervalent bonds exhibit greater orbital overlap compared to metallic bonds, resulting in steep and widely dispersed bands.[16, 23] The double layers of the La atoms are stacked by 6$_3$ spiral axes parallel to the c direction, giving rise to La-based zig-zag chains with the nearest La-La distance of 4.01 Å.

To gain a deeper understanding of the electronic structure of La$_3$ScBi$_5$, the projection of Bi $p$ orbitals onto the density functional theory band structure was analyzed according to the *4d* Wyckoff position, as shown in Fig. 1d. It is observed that the band near the Fermi level exhibits a significant contribution from 1D Bi chains. It is worth noting that the band projection between **Γ** and **A** reveals a steep slope, indicating a high Fermi velocity and suggesting excellent potential for thermal conductivity. This is primarily attributed to the 1D hypervalent Bi chains characteristics along the *c*-axis, where electron hopping is significantly greater than other directions, leading to more pronounced band dispersion. The partial density of states (DOS) incorporating SOC in Fig. 1e shows a small, broad peak near the Fermi level, predominantly arising from the total Bi $p$ orbitals, with notable hybridization from the La $d$ and Sc $d$ states. Taken together, the hypervalent Bi chains are essential in determining a range of physical properties, and additional experiments are required to delve into the thermal properties of La$_3$ScBi$_5$.

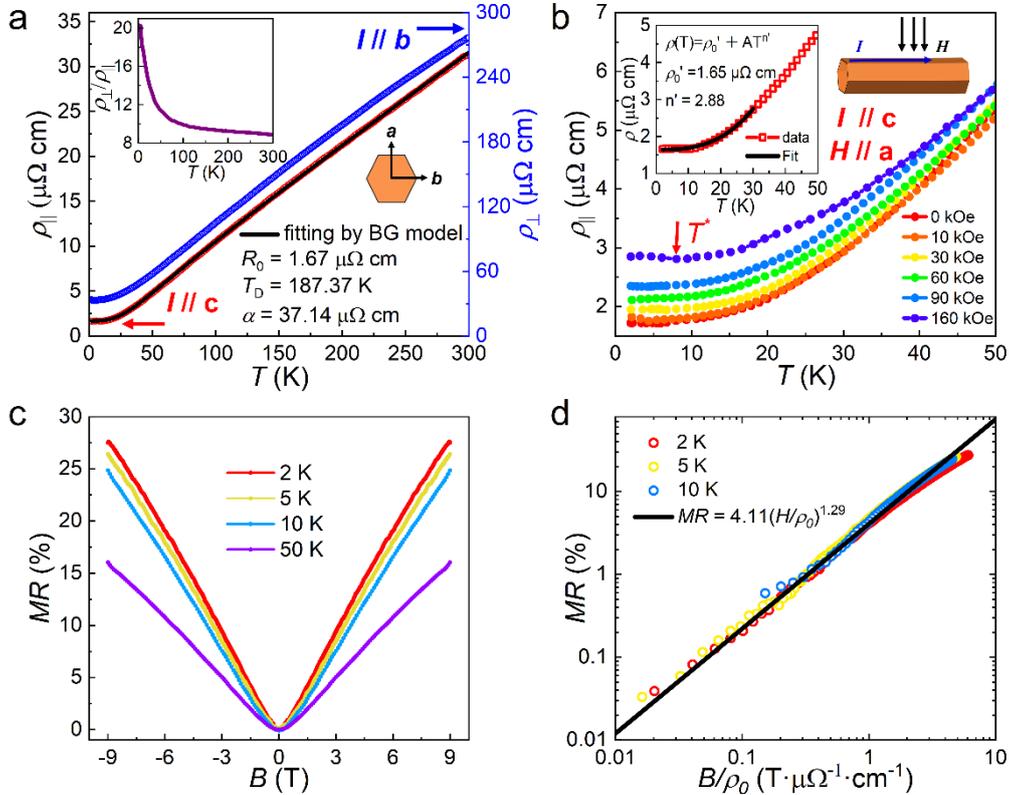

**Fig. 2 Charge transport of La$_3$ScBi$_5$ of $\rho_\parallel(T)$ (*I // c*). a** Temperature dependence of $\rho_\parallel$(*I // c*) and $\rho_\perp(T)$ (*I // b*) under zero field on sample S1. The solid line is the fit of data to the BG formula. Inset:

Temperature dependence of the radio of $\rho_\perp/\rho_\parallel$. **b** Temperature dependence of $\rho_\parallel(T)$ with $H // a$ ($I // c$) up to 16 T below 50 K. The red arrows denote the temperature $T^*$. Inset: $\rho_\parallel(T)$ below 50 K and fitting curve (solid line). **c** Field dependence of magnetoresistance at various temperatures. **d** Kohler plot: MR versus $B/\rho_\parallel(T, B = 0)$. The solid line is the fit to MR = $4.11(B/\rho_0)^{1.29}$.

## Quasi-linear Magnetoresistance and Nonlinear Hall Resistivity

Temperature-dependent resistivity of materials reflects the scattering of charge carriers by impurities and phonons at low and high temperatures, respectively. In La$_3$ScBi$_5$, the electrical resistivity, $\rho_\parallel(T)$ ($I // c$) and $\rho_\perp(T)$ ($I // b$), were investigated under a zero field as a function of temperature, as depicted in Fig. 2a. La$_3$ScBi$_5$ exhibits a metallic character with a residual resistivity ratio ($\rho_\parallel$, 300 K/$\rho_\parallel$, 2 K) of approximately 19. This *RRR* value, while not exceptionally larger, suggests the presence of significant disorder scattering. Despite the similar overall temperature dependencies of resistivity along both crystallographic axes, it was observed that $\rho_\perp$ is an order of magnitude higher than $\rho_\parallel$, yielding an anisotropy ratio ($\rho_\perp/\rho_\parallel$) of around 20 at 2 K. The high-temperature behavior of $\rho_\parallel(T)$ can be effectively described by the Bloch-Grüneisen (BG) formula [25]:

$$\rho(T)_{BG} = \rho_0 + \alpha \left(\frac{T}{T_D}\right)^n \int_0^{\frac{T_D}{T}} \frac{x^n dx}{(e^x - 1)(1 - e^{-x})} \quad (1)$$

Here, $\rho_0$ represents the residual resistivity, $T_D$ corresponds to the Debye temperature. By utilizing the optimal parameters of n = 3, with values of $\rho_0$ = 1.67 μΩ cm, $T_D$ = 187.37 K, and $\alpha$ = 34.17 μΩ cm/K, the BG formula provides a good fit for $\rho_\parallel$ in the temperature range of 2 to 300 K. The obtained $T_D$ value closely aligns with that derived from the specific heat of La$_3$ScBi$_5$ (Fig. S4, Supporting Information). The low-temperature (< 30 K) $\rho_\parallel$ follows a power law behavior $\rho(T) = \rho_0' + AT^{n'}$ with $\rho_0'$ = 1.65 μΩ cm and n$'$ = 2.88, as shown in the inset of Fig. 2b. Both values of n are nearly exactly 3, exhibiting a $T^3$ dependence, indicating that the primary interaction in our system is likely inter-band electron-phonon scattering[26, 27]. Similar analyses and interpretations have been applied to isomorphic compounds such as Sm$_3$ZrBi$_5$[16], as well as to ZrSiS[28] and TiS$_2$[29].

Figure 2b illustrates the temperature dependence of $\rho_\parallel(T)$ under various fields with the $H // a$ axis. In the absence of an external magnetic field, the resistivity decreases monotonically with a reduction in temperature. However, upon ramping the magnetic field to 16 T, a distinctive turn-on behavior is observed, leading to a resistivity minimum at $T^*$ before eventually plateauing. This magnetic-field-driven resistivity upturn followed by a resistivity plateau is interpreted as a potential transport hallmark of conducting surface states, attributed to the LL quantization of relativistic electrons.[30, 31] Such a turn-on behavior has been documented in topological materials such as PtTe$_2$[32] and WTe$_2$.[33] Another viable mechanism posits that the reentrant metallic state is a consequence of the scaling behavior in magnetoresistance, driven by the power-law dependence of the magnetic field and temperature.[34] The corresponding field-dependent MR ratio, $[\rho_\parallel(B)- \rho_\parallel(0\text{ T})]/\rho_\parallel(0\text{ T})$, at various temperatures is depicted in Fig. 2c. Notably, the MR value peaks at 28 % at 2 K and 9 T, displaying no saturation tendency. Particularly intriguing is the nearly linear dependence of MR on the magnetic field, most evident beyond the low-field regime. A similar quasi-linear positive MR phenomenon has been observed in HoAgGe, attributed to uncompensated charge carriers.[35] It's noteworthy to mention that this linear MR might also be a signature of charge density wave materials, including $TbTe_3$, $HoTe_3$,[36] and $LaAuSb_2$.[37] The existence of inhomogeneity and open Fermi surfaces or electron-phonon coupling along with a nested imperfect Fermi surface may be plausible mechanisms.[36, 38, 39]

To further research the underlying electronic structure influencing the MR behavior, we have conducted an analysis based on Kohler's rule in Fig. 2d, which dictates that the MR in a conventional metal obeys a scaling behavior of $\Delta\rho/\rho(0) = F[B/\rho(0)]$. The MR data at low temperatures exhibit a remarkable collapse onto a singular curve, which can be effectively approximated by the formula: MR = $4.11(B/\rho_0)^{1.29}$, indicating that $La_3ScBi_5$ does not conform to an electron-hole compensated system.[6, 40] The MR depicted in Fig. S5a (Supporting Information) was calculated using the WannierTools package, employing the Boltzmann transport equation and Wannier function techniques. The calculated $\rho_{\parallel}\tau$ demonstrates a nonsaturating dependence of $(B\tau)^{1.25}$, which is quite consistent with the experimental results of MR $\propto B^{1.29}$. The agreement between theoretical calculation and experiment suggests that the MR may be explained by the semiclassical theory, with no significant role from the Berry curvature aspect of topological Dirac points. The adherence to Kohler's rule serves as a pivotal indication; it underscores that the observed upturns and low-temperature plateaus are not indicative of a metal-insulator transition but rather stem from factors such as small residual resistivity, high mobilities, and a low charge-carrier density.[6, 33, 34]

Our investigation also systematically explored the transport properties along the c-crystallographic directions, unveiling anisotropic characteristics in the seemingly unsaturated MR. The $\rho_{\perp}(T)$ curves obtained under various magnetic fields are compiled in Fig. 3a, where the turn-on behavior is also observed at 3 T and above. The saturation value and the range of upturn amplify with the applied magnetic field strength, indicative of the positive MR effect. Figs. 3b and 3c illustrate the magnetic field dependence of MR at diverse temperatures and the field derivative of MR, $d$MR/$dB$, respectively. The quasi-linear behavior extends to notably low crossover fields $B^*$, beyond which MR transitions to a weak-field semiclassical quadratic dependence, which is consistent with the behavior of the theoretically calculated MR (Fig. S5b, Supporting Information). Here, the $B^*$ denotes the point where the fitting lines intersect. With increasing temperature, the field range exhibiting quasi-linear MR diminishes, and MR decreases, consistent with observations in *112*-type materials such as $CaMnBi_2$,[41] $YbMnBi_2$[42] and $LaAgBi_2$,[43] suggesting the possible existence of a nontrivial state. In the quantum limit where all carriers occupy solely the lowest LL, the observed $B^*$ corresponds to the quantum limit of $B^* = (1/2e\hbar v_F^2)(k_BT + E_F)^2$.[44, 45] As depicted in Fig. 3d, the experimental data for $B^*$ align well with the above equation, further indicating the possibility of nontrivial states in $La_3ScBi_5$. Moreover, with increasing temperature, the linear term coefficient $A_1$ is suppressed due to the temperature smearing of the LL splitting.[43] Future investigations, requiring more systematic Angle-Resolved Photoemission Spectroscopy (ARPES) studies, are necessary to probe the existence of Dirac cones in the electronic structure. Such the peculiar behavior of MR is predominantly governed by the multi-band features near the Fermi surface. Through the projection of individual atomic orbitals (Fig. S2, Supporting Information) and DOS analysis, it is evident that the Bi *p* orbitals exhibit substantial contributions, as do hybrids between other orbitals, which together have a comprehensive effect on MR.

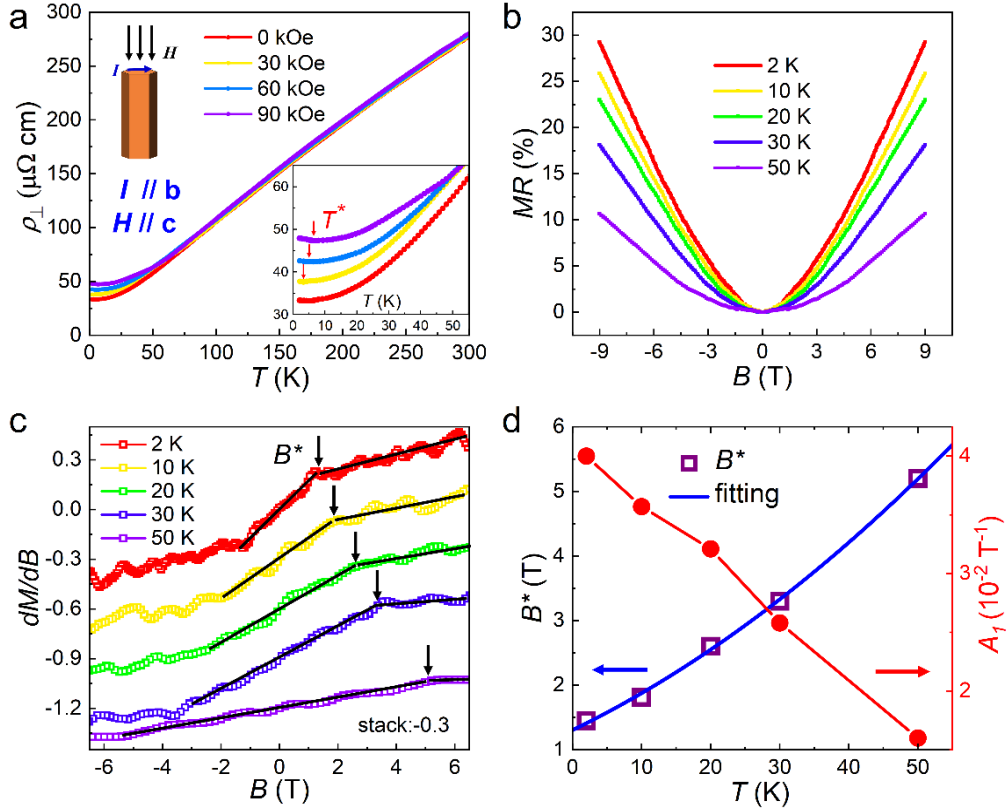

**Fig. 3 Charge transport of La$_3$ScBi$_5$ of $\rho_\perp(T)$ ($I // b$). a** Temperature dependence of $\rho_\perp(T)$ with $H // c$ ($I // b$) under various fields on sample S1. Inset: Enlarged view of $\rho_\perp(T)$ below 50 K. The red arrows denote the temperature $T^*$. **b** Field dependence of MR at various temperatures. **c** The field derivative of MR at various temperatures. In the high-field regime, fitting results using MR = $A_1B$ + $O(B^2)$ are represented by lines, while in the low-field region, MR = $A_2B$ is employed. **d** Temperature dependence of the critical field $B^*$ (Purple squares). The solid line depicts the fitting results of $B^*$, utilizing $B^* = (1/2e\hbar v_F^2)(k_BT + E_F)^2$. The red circle corresponds to the linear coefficient $A_1$ in the high-field MR.

To determine the mobility of the charge carriers, Figure 4a illustrates the field dependence of the Hall resistivity $\rho_{yx}$ at various temperatures, revealing a nonmonotonic behavior in La$_3$ScBi$_5$ below 30 K, indicative of its multiband nature. At high temperatures ($T \geq 30$ K), $\rho_{yx}$ remains positive and exhibits a linear relationship with the applied magnetic field, suggesting predominant transport by single-hole-type carriers. Fig. 4b presents the field dependence of Hall resistance considering a relaxation time of 0.17 ps. Remarkably, our calculation shows the same trend as the experimental measurements, which exhibit nonlinear behavior that demonstrates the multicarrier behavior. We apply Kohler's rule for Hall resistivity to La$_3$ScBi$_5$, demonstrating that the Hall resistivity scales with both magnetic field and temperature, akin to the longitudinal resistivity scaling under the relaxation time approximation. This theoretical framework was initially proposed by Zhang et al.,[46] and is further elaborated in Note 1 (Supporting Information). This observed scaling behavior shows reasonable agreement with theoretical calculations. Compared with the case of $\rho_{||}$, Kohler's rule is basically obeyed over a large temperature range for $\rho_{yx}$ (Fig. 4c). Hence, a multi-band model can be used to analyze the obtained results:[47, 48]

$$\sigma_{xy}(B) = \frac{\rho_{yx}}{\rho_{||}^2 + \rho_{xy}^2} = eB \sum_i^N \frac{n_i \mu_i^2}{1 + \mu_i^2 B^2}, \quad (2)$$

$$\sigma_{xx}(B) = \frac{\rho_{||}}{\rho_{||}^2 + \rho_{yx}^2} = e \sum_i^N \frac{n_i \mu_i}{1 + \mu_i^2 B^2}. \quad (3)$$

A typical fitting for $T = 2$ K is presented in Fig. 4d, yielding carrier densities of $n_{e1} = 1.6 \times 10^{21}$ cm$^{-3}$, $n_{e2} = 6.3 \times 10^{19}$ cm$^{-3}$, $n_{h1} = 1.9 \times 10^{19}$ cm$^{-3}$, and $n_{h2} = 5.1 \times 10^{18}$ cm$^{-3}$, along with mobility values of $m_{e1} = 139$ cm$^2$ V$^{-1}$ s$^{-1}$, $m_{e2} = 1184$ cm$^2$ V$^{-1}$ s$^{-1}$, $m_{h1} = 893$ cm$^2$ V$^{-1}$ s$^{-1}$, and $m_{h2} = 4276$ cm$^2$ V$^{-1}$ s$^{-1}$. It is noteworthy that an excessive number of fitting parameters may lead to significant errors, potentially failing to accurately reflect the physical essence of the multi-band system[47]. For the multi-band system of La$_3$ScBi$_5$, fitting the multi-band model with smaller $N$ values is particularly challenging, possibly due to the lack of suitable initial values. Similar limitations are observed in other multiband systems, such as the pyrite-type bismuthide PtBi$_2$[49] and the 3D Weyl semimetal TaAs[50].

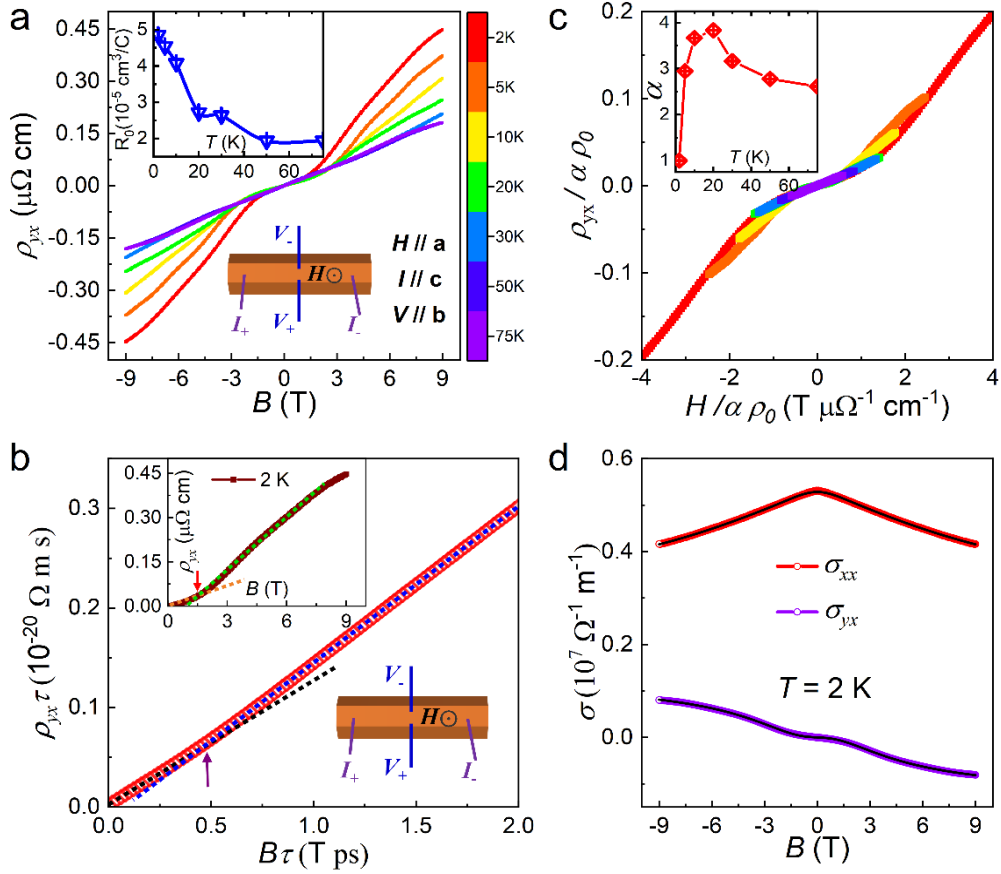

**Fig. 4 Hall effect of La$_3$ScBi$_5$. a** Field dependence of Hall resistivity $\rho_{yx}(B)$ at various temperatures. Inset: The temperature dependence of $R_0(T)$. A diagram of the Hall bar is shown, where I$^+$ and I$^-$ are a pair of current leads, and V$^+$ and V$^-$ are a pair of voltage leads, with $H$ // a, $I$ // c, and $V$ // b. **b** The calculated field dependence of Hall resistivity at $T$ = 2 K. Inset: The experimental field dependence of Hall resistivity at $T$ = 2 K. **c** $\rho_{yx}\alpha\rho_0$ versus $H/\alpha\rho_0$. $\rho_0$ is derived from $\rho_{||}(T, H=0)$ data. Inset: The temperature dependence of $\alpha$. **d** Conductivity and Hall conductivity of La$_3$ScBi$_5$ at $T$ = 2 K, the black line represents the fitted line using the multi-band model.

## Quantum Oscillations and Fermi Surface Properties

The quantitative analyses of dHvA oscillations reveal the nature of electrons and serve as a powerful tool in studying topological physics. Figure 5a shows the beautiful dHvA oscillations that persist to about 10 K for sample S1 for field orientation $H // a$ up to 9 T. The oscillatory components $\Delta M$, obtained by subtracting a smooth background, are plotted against $1/B$ in Fig. 5b. From the fast Fourier transform (FFT) analyses of the oscillatory components $\Delta M$, three principal frequencies ($F_\alpha$ = 33.8 T, $F_{\alpha'}$ = 56.1 T, and $F_\beta$ = 90.5 T) are derived (Fig. 5c). The $F_\alpha$ component exhibits a significantly greater oscillation amplitude compared to $F_{\alpha'}$ and $F_\beta$ components, making the $F_\alpha$ frequency the focal point of research interest. Similar multifrequency oscillations have also been observed in the isostructural compound $La_3MgBi_5$,[18, 19] which is suggestive of a 3D-like electronic structure. Note that similar dHvA measurements were observed on a crystal from the same batch as the sample S1(Fig. S7, Supporting Information). The oscillation frequency is directly linked to the cross-sectional area of the Fermi pockets normal to the magnetic field $A_F$ through the Onsager relation $F = (\Phi_0/2\pi^2)A_F$. Specifically, the $A_F$ associated with the main peak is estimated to be $3.23 \times 10^{-3}$ Å$^{-2}$ for the $\alpha$ bands, yielding the Fermi vector $k_\alpha = 3.20 \times 10^{-2}$ Å$^{-1}$. The corresponding parameters for three different pockets are listed in Table SIV (Supporting Information). It is noteworthy that distinct and varying quantum oscillations were observed in the $H // b$ orientation (Fig. S6, Supporting Information), indicating pronounced in-plane anisotropy within the $ab$-plane of $La_3ScBi_5$.

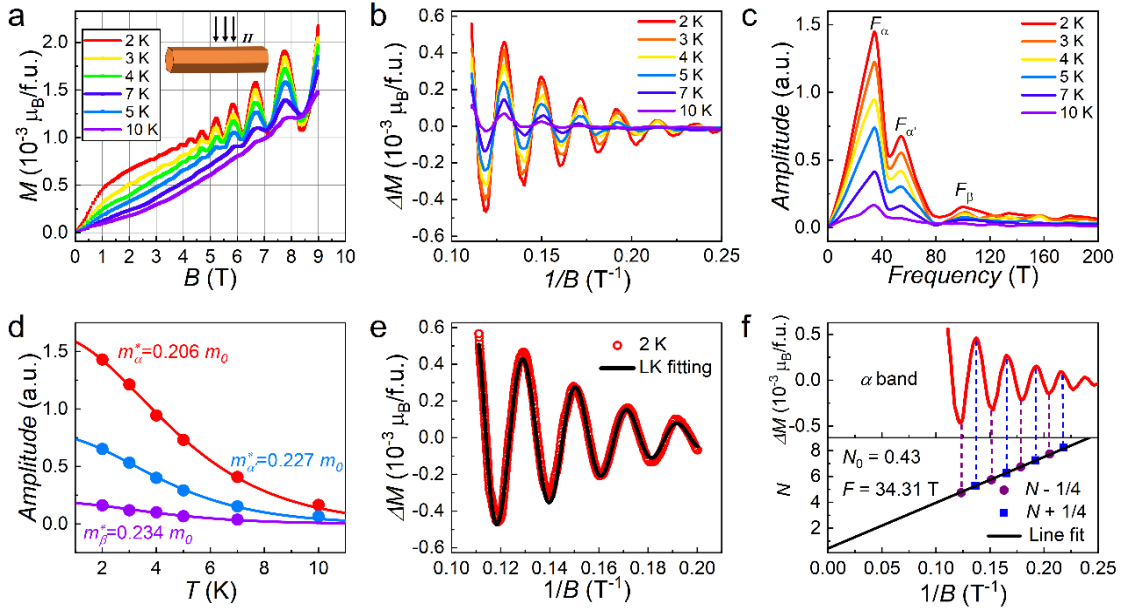

**Fig. 5 Analyses of the de Haas-van Alphen (dHvA) oscillations for $H // a$ in $La_3ScBi_5$. a** Isothermal magnetization at different temperatures with the magnetic field parallel to the a-axis on sample S1. **b** The oscillatory components of magnetization $\Delta M$. **c** The corresponding FFT spectrum of the oscillatory component of the dHvA oscillations at various temperatures. **d** The fits of the FFT amplitudes of $F_\alpha$, $F_{\alpha'}$, and $F_\beta$ to the temperature damping factor $R_T$ by the LK formula. **e** Multiband LK fit of the oscillation components of the dHvA oscillations at 2 K. The black lines show the fits of the oscillation pattern to the generalized three-band LK formula for three different frequency oscillations. **f** The LL index fan diagram for $\alpha$ band derived from oscillatory magnetization at 2 K.

The oscillatory magnetization of a Dirac system is generally described by the Lifshitz-Kosevich (LK) formula,[51] accounting for the Berry phase:[52]

$$\Delta M \propto -B^\lambda R_T R_D R_S \sin\left[2\pi\left(\frac{F}{B} - \gamma - \delta\right)\right] \quad (4)$$

where $R_T = (KT\mu/B)/\sinh(KT\mu/B)$, $K = 2\pi^2 k_B m_e/(\hbar e) \approx 14.69$ T/K, $R_D = \exp(-KT_D\mu/B)$. $\mu$ represents the ratio of effective cyclotron mass to free electron mass, while $R_S$ denotes the spin reduction factor arising from Zeeman splitting. The oscillation of $\Delta M$ is delineated by the sine term with the phase factor $1/2 - \varphi_B/2\pi - \delta$. The phase shift $\delta$ adopts values of 0 and $\pm 1/8$ for the 2D and 3D Fermi surfaces, respectively. Utilizing the LK formula, the effective mass $m^*$ can be derived by fitting the temperature-dependent oscillation amplitude to the thermal damping factor $R_T$, as depicted in Fig. 5d. Here, $B$ denotes the inverse average of the field window employed for Fourier analysis,[53] given by $B_{ave} = [(1/B_{max} + 1/B_{min})/2] = 6.43$ T with $B_{max} = 9$ T and $B_{min} = 5$ T, leading to calculated $m^*/m_0$ of 0.206, 0.227 and 0.234 for $\alpha$, $\alpha'$ and $\beta$ pockets, respectively. The small values of $m^*$ imply the presence of relativistic charge carriers. Given that the Fermi surface of La$_3$ScBi$_5$ is of a strong 3D character, we have adopted the oscillation pattern at 2 K using the multiband LK formula with fixed frequency and fixed effective mass (Fig. 5e), yielding the Dingle temperature $T_D = 4.56$ K for the $F_\alpha$ band, which corresponds to the quantum relaxation time $\tau_q = \hbar/(2\pi k_B T_D) = 2.67 \times 10^{-13}$ s, and quantum mobility $\mu_q = e\tau_q/m^*_\alpha = 2726$ cm$^2$/ V$^{-1}$ s$^{-1}$. From the LK fit of oscillations with $F_\alpha = 33.8$ T, a phase factor of $-\gamma - \delta = 1.97$ is obtained, yielding the Berry phase $\varphi_B$ is determined to be $0.69\pi$ ($\delta = -1/8$) or $1.19\pi$ ($\delta = 1/8$). Based on the calculated Fermi surface presented below, with $\varphi_B = 0.69\pi$ ($\delta = -1/8$), the oscillation of the probe $F_\alpha$ corresponds to the minimum cross-section. The Dingle temperature, relation time, quantum mobility, and Berry phase obtained by LK fitting for $F_{\alpha'}$ and $F_\beta$ are summarized in Table SIV (Supporting Information). Certainly, the fitting performed below 5 T is based on a limited number of data points, which may introduce greater uncertainty for Berry phases with multiple LK fits.

Given that $dM/dB$ is directly proportional to the density of states at the Fermi level,[54,55] we can associate the minimum of $\Delta M$ with $N-1/4$, where $N$ represents the LL index. However, due to the Landau indices of $F_\beta$ being far from the quantum limit, there is notable uncertainty in determining the intercept. As depicted in Fig. 5f, the LL indices $N$ for $\alpha$ bands are plotted against $1/B$. The solid lines depict linear fits based on the Lifshitz-Onsager quantization criterion $N = F/B + \varphi_B/2\pi - \delta$. The slope of the linear fits yields a value of 34.31 T, consistent with the FFT derived result, thereby affirming the reliability of the linear fit within the LL fan diagram. The intercept extracted from linear extrapolation is $0.430 \pm 0.018$, corresponding to a Berry phase of $0.72\pi$ ($\delta = -1/8$), aligning with the findings obtained from LK fitting. While a finite Berry phase is generally considered indicative of topologically nontrivial bands, it may not reflect the intrinsic properties of a 3D-Dirac Fermi surface, but could simply be a consequence of time-reversal symmetry.[56] However, the phase shift can be affected by spin splitting and g-factor, making it challenging to determine the topological properties of the band from magnetic transport measurements alone. The similar deviation in the intercept was observed in the study of PtBi$_2$[57], which was attributed to the Fermi level not intersecting the degeneracy point.

In a manner analogous to the in-plane cyclotron motions observed for $H$ // a, electrons engaged in interlayer cyclotron motions under fields along the c-axis exhibit quantum oscillation behavior. Figs. S8a and S8b (Supporting Information) depict dHvA oscillations of $H$ // c overlaid on a diamagnetic background in sample S1 and their respective oscillatory components. In contrast to the dHvA oscillations for $H$ // a, those for $H$ // c consist of more frequencies. FFT analyses (Fig. S8c, Supporting Information) reveal a predominantly low frequency (106.8 T) and three high frequencies (181.8, 215.2, and 249.1 T), and this phenomenon is also reproduced for samples S2 and S3 (Fig. S9, Supporting Information). The electron cyclotron masses for $H$ // c range slightly higher, approximately between 0.366 and 0.529 m$_0$ across all probed oscillation frequencies. This anisotropic effective mass is consistent with the quasi-1D structure of La$_3$ScBi$_5$. The lower single-frequency component, extracted by isolating the higher-frequency component, can be analyzed using the widely adopted LL fan diagram method, as demonstrated in Fig. S8d (Supporting Information). The intercept derived from the linear fit in the fan diagram yields 0.234 ± 0.092 for $F_\gamma$, with uncertainties suggesting $\delta$ values of 1/8 or -1/8, resulting in an uncertain phase of 0.71$\pi$ or 0.21$\pi$. Additionally, the fitted oscillation frequency for $F_\gamma$ is 108.09 T, closely aligning with FFT results. Noted that significant fitting errors limit the precision of determining $\varphi_B$ for γ band. Furthermore, accurately determining the LL index field for each frequency remains challenging in the case of multifrequency oscillations for $F_\eta$.[54, 58] Nonetheless, we did not detect Shubnikov-de Hass (SdH) oscillations either in the MR or in the Hall resistivity up to 9 T, which warrants further measurements at higher magnetic fields and lower temperatures. It is conceivable that SdH oscillations come from the oscillating scattering rate and can thus be complicated by the detailed scattering processes while the dHvA effect is caused directly by the free energy oscillations of a system.[59, 60]

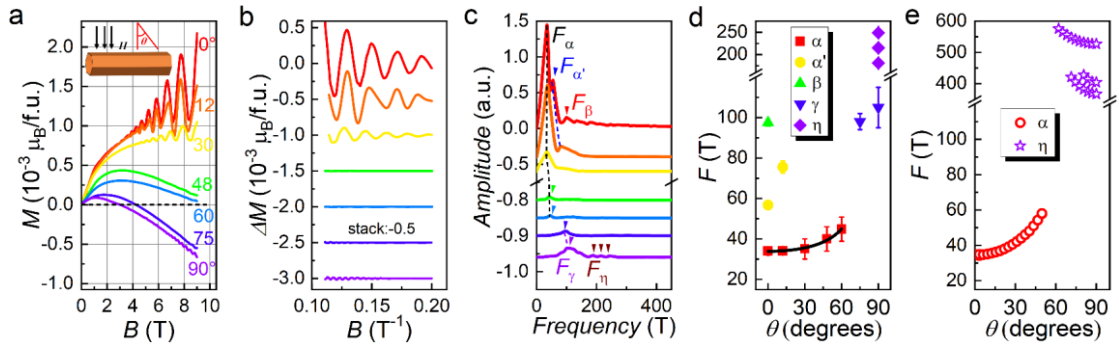

**Fig. 6 Fermi surface morphology of La$_3$ScBi$_5$. a** Isothermal magnetization of La$_3$ScBi$_5$ at various angles for $T$ = 2 K on sample S1. **b** Oscillatory components of magnetization at 2 K under different magnetic field orientations. **c** The corresponding FFT spectrum of the oscillatory component of the dHvA oscillations for various angles. **d** Angular dependence of dHvA oscillation frequencies, with error bars defined as half the width at the half-height of FFT peaks. Black lines represent fits to $F = F_{3D} + F_{2D}/\cos\theta$ for $F_\alpha$. **e** Angular dependence of calculated dHvA oscillation frequencies below 600 T. $F_\alpha$ contributed from $e$ pocket is indicated by red circles while $F_\eta$ contributed from $h$ pocket is denoted with purple stars.

The anisotropic characteristics of the Fermi surface of La$_3$ScBi$_5$ are further elucidated through an investigation of $\Delta M$ versus $1/B$ and corresponding dHvA oscillation frequencies at 2 K across various magnetic field orientations. Figure 6a illustrates angle-dependent quantum oscillation as the magnetic field rotates from the [100] to [001] direction. Figure 6b reveals the oscillatory component after subtracting the background, demonstrating a distinct evolution of the angle $\theta$, defined as the angle between the field direction and [100] direction. The oscillation amplitude diminishes progressively with increasing the tilt angle. Additionally, there is one more frequency $F_\gamma$ detected at $\theta \geq 75°$ in the FFT spectra (Fig. 6c). These angular dependencies of dHvA oscillation frequencies clearly indicate the complex 3D nature of the Fermi surface morphology in La$_3$ScBi$_5$ despite its quasi-1D structure. In Fig. 6d, we summarize the angle dependence of each frequency. The oscillation frequencies $F_\alpha$ exhibits a weak angular dependence at lower angles and can be fitted with the formula $F = F_{3D} + F_{2D}/\cos\theta$, where $F_{2D}$ and $F_{3D}$ represent the contributions from 2D and 3D components. The relative weight between 2D and 3D components derived from the fit, $F_{2D}/F_{3D} \approx 0.6$, suggests dimensionality between 2D and 3D. Moreover, Figure S10 presents the Fermi surfaces incorporating SOC, and the complexity of these Fermi surfaces leads to an angular dependence in dHvA oscillation frequencies. Utilizing the SKEAF package,[74] we calculated the angular dependence of the dHvA oscillations, as depicted in Fig. 6e, which shows general agreement with the experimental results. These calculations enable us to identify the origins of the dHvA frequencies observed in experiments. The Fermi surface is intricate, revealing in particular one electron pocket and one hole pocket, which may correspond to the $\alpha$ and $\eta$ branches in the angular dependence of dHvA oscillation frequencies, respectively (Fig. S10, Supporting Information). In conjunction with the projections of each atom, it is evident that the Bi *p* orbitals play a crucial role in this electron pocket.

**Paramagnetic Singularity in Magnetization.**

The magnetization of La$_3$ScBi$_5$ exhibits pronounced anisotropy with respect to the field orientations, as depicted in Fig. 7a. Along the a-axis ($\chi_a$) and c-axis ($\chi_c$), the temperature-dependent magnetic susceptibility shows distinct behaviors. $\chi_a$ manifests paramagnetic, nearly independent of temperature, with a broad shoulder around 75 K and a notable upturn below 15 K. These features persist as the applied field increases, ruling out magnetic impurity effects. Additionally, Figure 7b reveals the absence of anomalies near 75 K in $d\chi/dt$, providing further evidence of the intrinsic nature of this magnetic response in La$_3$ScBi$_5$. Similar *T*-independent susceptibility behaviors have been observed in bismuthides like Bi$_2$Se$_3$[61] and PtBi$_2$.[62, 63] This behavior can be attributed to potential contributions from Pauli paramagnetism, van Vleck paramagnetism, Landau diamagnetism, and core diamagnetism.[59, 61, 62]

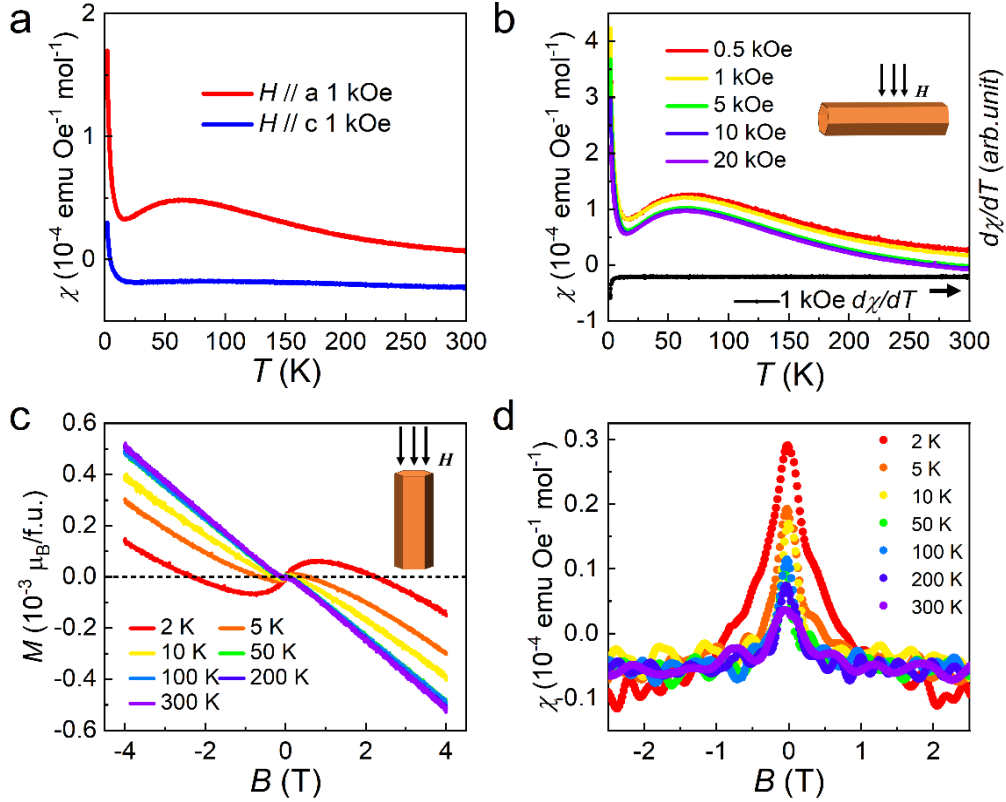

**Fig. 7 Magnetism of La$_3$ScBi$_5$. a** Temperature dependence of the magnetic susceptibilities measured at $H$ = 1 kOe on sample S1. **b** $\chi(T)$ curves under various magnetic fields for $H \parallel$ a. **c** Low field region of the magnetic moment vs. $B$ curves at various temperatures for $H \parallel$ c. The measurements are performed sequentially from low to high temperature, spanning from 2 K to 300 K. **d** Magnetic susceptibility $\chi(=dM/dB)$ is calculated by differentiating the magnetic moment, plotted as a function of the magnetic field (additional details are provided in Fig. S13, Supporting Information).

In contrast, the interplane $\chi_c$ is diamagnetic and undergoes a rapid increase at low temperatures. Specifically, the low-field region of $M(B)$ curves in Fig. 7c shows a significant paramagnetic contribution near the zero field, as also observed in Fig. S9 for the same sample batch. The $\chi(B)$ curves sharply rise above the diamagnetic 'floor' within a narrow field range and approach $\chi(0)$ in a straight line (Fig. 7d). The cusp observed in the low-field region are robust and prominent, with singular field dependence persisting up to the maximum measured temperature of 300 K. Additionally, there is a noticeable directional dependence, leading us to speculate that these features may be an inherent property of La$_3$ScBi$_5$ crystals, rather than being attributed to magnetic impurities in the diamagnetic matrix. With increasing temperature, the cusp exhibits a tendency to be inhibited, which may may be due to aging effect (Fig. S12, Supporting Information) or insufficient measurement accuracy, but also need further systematic research. Such cusp-like paramagnetic susceptibility be traced most naturally to the helical spin texture of topological electrons on the surface, a signature observed across the family of three-dimensional topological materials.[64, 65] The proposed mechanism suggests that the electron spins at the Dirac point lack a specific orientation. As long as the Dirac cone is not fragmented, these electron spins can align freely along the external magnetic field. In the low-field region, such freely oriented spins are expected to generate paramagnetic singularities in the $M$(B) curve.[64] More experiments, such as ARPES and scanning tunneling microscopy measurements, are highly desirable to probe the nontrivial surface and bulk states in La$_3$ScBi$_5$.

## DISCUSSION AND CONCLUSION

We have conducted comprehensive investigations into the physical properties of the quasi-one-dimensional single crystal La$_3$ScBi$_5$, belonging to the *Ln$_3$MX$_5$* family (*Ln* = rare-earth; *M* = transition metals; *X* = As, Sb, Bi). Density functional theory calculations reveal a complex and non-trivial electronic structure with SOC. Transport measurements reveal quasi-linear anisotropic magnetoresistance, with the MR isotherm following the Kohler scaling law with an index of m = 1.29, deviating from ideal electron-hole compensation. Nonlinear Hall resistivity further confirms the multi-band nature of electrical conductivity in La$_3$ScBi$_5$, with moderate charge compensation. The theoretical calculations agree with the experimental results of the behavior, demonstrating that both complex magnetoresistance and Hall resistivity behaviors may be accounted for by the semiclassical Boltzmann transportation theory.

Our magnetism studies demonstrate anisotropic magnetic susceptibility and detect dHvA oscillations for magnetic fields aligned parallel to both the a and c axes. Angle-dependent dHvA oscillations are essential to gain deeper insights into its fermiology and potential topological nature of the system. Notably, a robust and prominent paramagnetic singularity may underscore the non-trivial nature of the surface states, marking the groundbreaking observation within the *Ln$_3$MX$_5$* family. This research offers valuable insights into the behavior of similar quasi-one-dimensional systems containing heavy elements. The rod-like La$_3$ScBi$_5$, suitable for mechanical exfoliation, presents an ideal platform for device applications and topological physics.

# METHODS

*Crystal Growth:* Large single crystals of La$_3$ScBi$_5$ were grown using the self-flux method with excess Sc and Bi as flux. La (lump), Sc (piece), and Bi (pill) were initially mixed in a molar ratio of 1: 2: 5, loaded into an alumina crucible, and sealed into an evacuated quartz tube. The temperature was gradually raised to 1000 °C and maintained for 20 h, followed by slow cooling to 700 °C at a rate of 1 °C/h to facilitate single crystal growth. Subsequently, the excess flux was separated from the resulting rod-shaped La$_3$ScBi$_5$ crystals using a centrifuge. It's worth noting that prolonged exposure to air or contact with alcohol would cause decomposition.

*Bulk Characterization:* Single crystal x-ray diffraction (SCXRD) for La$_3$ScBi$_5$ was conducted on a Bruker D8 Venture diffractometer at room temperature using Mo Kα radiation (λ = 0.71073 Å). The collected data were refined by the least-square method of $F^2$ using SHELXL-2018/3.[66] The diffraction peaks for a single-crystalline surface were obtained via a Bruker D2 phaser XRD detector using Cu Kα1 radiation (λ = 1.54184 Å). Element analysis was performed using energy dispersive x-ray (EDX) in a Hitachi S-4800 scanning electron microscopy (SEM) with an accelerated voltage of 15 kV. Characteristic details can be shown in Figs. S1(a-f).

*Transport and Magnetism Measurements:* Electronic resistivity measurements up to 16 T were conducted on polished crystals to remove Bi flux droplets on the surface using a Physical Properties Measurement System (PPMS, Quantum Design). Magnetic susceptibility and isothermal magnetizations of La$_3$ScBi$_5$ were performed in a 9 T PPMS in vibrating sample magnetometer (VSM) mode. Heat capacity was measured by the thermal relaxation method in the PPMS, with thermal contact achieved using Apezion N-grease. Three different single crystal samples named S1, S2, and S3 were used for measurements, with mass bits of 32 mg, 80 mg, and 190 mg respectively.

*Electronic Structure and Transport Calculations:* Theoretical calculations were performed using density functional theory (DFT) as implemented in the Vienna ab initio Simulation Package (VASP)[67, 68] with the generalized gradient approximation (GGA) of Perdew-Burke-Ernzerhof (PBE) exchange-correlation potential.[69] A plane wave basis with a kinetic energy cutoff of 500 eV was employed. Brillouin zone sampling was conducted using a 4×4×6 k-point mesh, with Gaussian smearing of 0.05 eV applied around the Fermi surface. Structural optimization was carried out with both cell parameters and internal atomic positions allowed to relax until the forces on all atoms were less than $10^{-7}$ eV Å$^{-1}$. Spin-orbit coupling effects were included in the electronic property calculations. MR was computed using the WannierTools package,[70] employing the Boltzmann transport equation[71, 72] and Wannier function techniques.[73] These calculations utilized a 100×100×100 k-point mesh, with the assumption of identical relaxation times (τ) for charge carriers. The angular dependence of dHvA oscillation frequencies was calculated using the SKEAF package.[74]

## ACKNOWLEDGEMENTS

We acknowledge very invaluable advice with Shaokui Su and Huifen Ren regarding transport measurements. This work was supported by the National Key R&D Program of China (Grant No. 2024YFA1408400, 2023YFA1607400, 2022YFA1403800, 2021YFA1400401), the National Natural Science Foundation of China (Grant No.12274436, 11925408, 11921004, 52271238, U22A6005), the Science Center of the National Natural Science Foundation of China (Grant No. 12188101), the Strategic Priority Research Program of the Chinese Academy of Sciences (Grant No. XDB33010000), the Center for Materials Genome, and the SynergeticExtreme Condition User Facility (SECUF)

## AUTHOR CONTRIBUTIONS

Z. C. X. and Y. Y. contributed equally to this work. Z. C. X. carried out the preparation of the samples and conducted the structure, magnetic, and transport measurements. Y. Y. and Z. H. L. performed the ab initio calculation. X. M. Z., Y. G. S., and Q. S. W. conceived the idea and supervised the project. All authors discussed the results and contributed to writing the manuscript.

# Supporting Information:

# Quasi-linear magnetoresistance and paramagnetic singularity in Hypervalent Bismuthide

Zhongchen Xu[1,2,3], Yi Yan[3,4], Zhihao Liu[3,4], Jie Pang[2,3], Guohao Dong[2,3], Xiutong Deng[2,3], Shengnan Zhang[5], Xianmin Zhang[1], Youguo Shi[2,3,6], and Quansheng Wu[3,4,6]

[1]*Key Laboratory for Anisotropy and Texture of Materials (Ministry of Education), School of Material Science and Engineering, Northeastern University, Shenyang 110819, China*

[2]*Center of Materials Science and Optoelectronics Engineering, University of Chinese Academy of Sciences, Beijing 100190, China*

[3]*Beijing National Laboratory for Condensed Matter Physics and Institute of Physics, Chinese Academy of Sciences, Beijing 100190, China*

[4]*School of Physical Sciences, University of Chinese Academy of Sciences, Beijing 100190, China*

[5]*Beijing Polytechnic College, Beijing, 100042, China*

[6]*Songshan Lake Materials Laboratory, Dongguan, Guangdong 523808, China*

# Supporting Information Contents

## I. *Single crystal X-ray Diffraction information:*

1. Table SI shows crystallographic and structure refinement data of $La_3ScBi_5$.

2. Table SII shows atomic coordinates and equivalent isotropic thermal parameters of $La_3ScBi_5$.

3. Table SIII shows anisotropic atomic displacement parameters ($Å^2$) for $La_3ScBi_5$.

## II. *Characteristic parameters derived from the dHvA oscillations:*

4. Table SIV shows the parameters of different bands investigated through dHvA oscillations

## III. *Sample structure characterization :*

5. Figure S1 shows characterization of $La_3ScBi_5$.

## IV. *Additional text:*

6. Note 1 shows the Kholer's rule of Hall resistivity.

## V. *Additional experimental and computational data:*

7. Figure S2 shows the projection band structure considering SOC of all atoms.

8. Figure S3 shows the electronic band structures of $La_3ScBi_5$ using the DFT method and the DFT+U method.

9. Figure S4 shows the heat capacity $C_p$ for $La_3ScBi_5$, which was well-fitted by the Debye-Einstein model.

10. Figure S5 shows the calculated magnetoresistance.

11. Figure S6 shows the isothermal magnetization of $La_3ScBi_5$ measured at 2 K in *ab*-plane.

12. Figure S7 shows analyses of the dHvA oscillations with *H* // a at *T* = 2 K for SM2, SM3.

13. Figure S8 shows analyses of the dHvA oscillations for *H* // *c* in $La_3ScBi_5$.

14. Figure S9 shows analyses of the dHvA oscillations with *H* // c at *T* = 2 K for SM2, SM3.

15. Figure S10 shows the Fermi surfaces in the presence of spin-orbit coupling.

16. Figure S11 shows the inverse susceptibility of $La_3ScBi_5$ under $H$ // a and $H$ // c ($H$ = 1 kOe) conditions.
17. Figure S12 shows the susceptibility cusp for a $La_3ScBi_5$ crystal on sample S1 at 2 K measured an hour and a day after the crystal growth.
18. Figure S13 shows the paramagnetic susceptibility cusp rides on a temperature dependent diamagnetic background

# I. Single crystal X-ray Diffraction information

**Table SI.** Data collection and structure refinement for $La_3ScBi_5$.

| Chemical formula | **$La_3ScBi_5$** |
|---|---|
| Temperature | 274(2) K |
| Formula weight | 1506.60 g/mol |
| Radiation | Mo $K\alpha$ 0.71073 Å |
| Crystal system | Hexagonal |
| Space group | $P6_3/mcm$ |
| Unit-cell dimensions | $a$ = 9.746(3) Å |
|  | $b$ = 9.746(3) Å |
|  | $c$ = 6.561(3) Å |
| Volume | 539.7(5) Å$^3$ |
| Z | 2 |
| Density (calculated) | 9.271 g/cm$^3$ |
| Absorption coefficient | 93.300 mm$^{-1}$ |
| F(000) | 1214 |
| Θ range for data collection | 2.41 to 33.14° |
| Index ranges | -14<=h<=14, |
|  | -14<=k<=13, |
|  | -10<=l<=10 |
| Independent reflections | 400 [$R_{(int)}$ = 0.0995] |
| Structure solution program | SHELXT 2018/2 (Sheldrick, 2018) |
| Refinement method | Full-matrix least-squares on $F^2$ |
| Refinement program | SHELXL-2018/3 (Sheldrick, 2018) |
| Function minimized | $\Sigma w (F_o^2 - F_c^2)^2$ |
| Data/restraints / parameters | 400 / 0 / 14 |
| Goodness-of-fit on $F^2$ | 1.102 |
| Final R indices | 360 data; I>2σ(I) |
|  | R1 = 0.0333, wR2 = 0.0905 |
|  | all data |
|  | R1 = 0.0395, wR2 = 0.0941 |
| Weighting scheme | w=1/[σ$^2$($F_o^2$) + (0.0530P)$^2$] |
|  | where P=($F_o^2$+2$F_c^2$)/3 |

Table SII. Crystallographic data of La$_3$ScBi$_5$.

| Atom | Wyckoff | Symmetry | x | y | z | Occup[a] | $U_{eq}$[b] |
|------|---------|----------|---------|---------|---------|------|-------|
| La   | 6g      | m2m      | 0.61808 | 0.61808 | 0.75000 | 1.000 | 0.014 |
| Sc   | 2b      | -3. m    | 1.00000 | 1.00000 | 1.00000 | 1.000 | 0.012 |
| Bi1  | 4d      | 3. 2     | 0.33333 | 0.66667 | 0.50000 | 1.000 | 0.014 |
| Bi2  | 6g      | m2m      | 0.73964 | 1.00000 | 0.75000 | 1.000 | 0.013 |

[a] *Occup*: Occupancy.
[b] $U_{eq}$: equivalent isotropic thermal parameter.

Table SIII. Anisotropic atomic displacement parameters (Å$^2$) for La$_3$ScBi$_5$.

| Label   | $U_{11}$    | $U_{22}$    | $U_{33}$    | $U_{23}$ | $U_{13}$ | $U_{12}$     |
|---------|-------------|-------------|-------------|----------|----------|--------------|
| Bi (01) | 0.0125(3)   | 0.0125(3)   | 0.0158(4)   | 0        | 0        | 0.00623(15)  |
| Bi (02) | 0.0110(3)   | 0.0123(4)   | 0.0162(4)   | 0        | 0        | 0.00616(18)  |
| La (03) | 0.0121(4)   | 0.0121(4)   | 0.0165(5)   | 0        | 0        | 0.0048(4)    |
| Sc (04) | 0.0113(16)  | 0.0113(16)  | 0.0014(3)   | 0        | 0        | 0.0057(8)    |

The anisotropic atomic displacement factor exponent takes the form: $-2\pi^2[\ h^2\ a^{*2}\ U_{11} + ... + 2hka^*b^*U_{12}]$

## II. Characteristic parameters derived from the de Haas–van Alphen oscillations.

**Table SIV.** The oscillation frequency ($F$), cross-section area ($A_F$), Dingle temperature ($T_D$), effective mass ($m^*/m_0$), quantum relaxation time ($\tau_q$), quantum mobility ($\mu_q$), and Berry phase ($\varphi_B$) of different bands investigated through dHvA oscillations.

| | $F$ (T) | $A_F$ (nm$^{-2}$) | $m^*/m_0$ | $T_D$ (K) | $\tau_q$ (ps) | $\mu_q$ (cm$^2$/V s) | $\varphi_B$ $\delta=-1/8$ | $\delta=0$ | $\delta=1/8$ |
|---|---|---|---|---|---|---|---|---|---|
| $\alpha$ band | 33.8 | 0.323 | 0.206 | 4.56 | 2.67×10$^{-13}$ | 2726 | 0.69π | / | / |
| $\alpha'$ band | 56.1 | 0.536 | 0.227 | 6.45 | 1.88×10$^{-13}$ | 1927 | 0.43π | 0.18π | 1.73π |
| $\beta$ band | 90.5 | 0.864 | 0.234 | 11.35 | 1.07×10$^{-13}$ | 1095 | 1.35π | 1.1π | 0.65π |
| $\gamma$ band | 106.8 | 1.020 | 0.366 | / | / | / | 0.71π | 0.46π | 0.21π |
| $\eta$ band | 181.8 | 1.735 | 0.408 | / | / | / | / | / | / |
| | 215.2 | 2.054 | 0.492 | / | / | / | / | / | / |
| | 249.1 | 2.378 | 0.529 | / | / | / | / | / | / |

III. Sample structure characterization

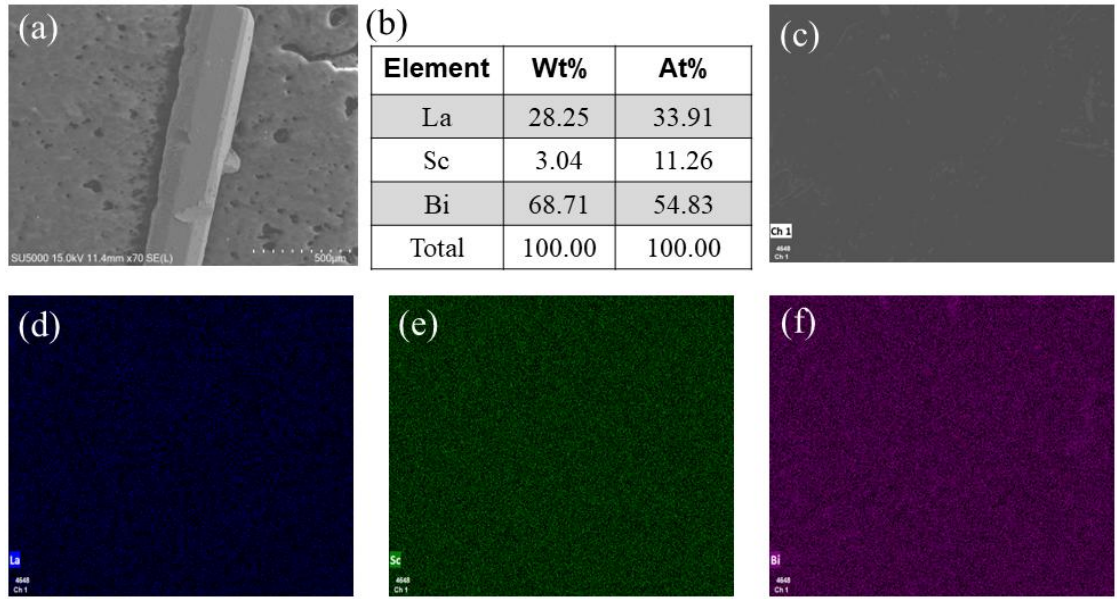

**Fig. S1 Characterization of La$_3$ScBi$_5$.** (**a**) SEM image of crystallographic morphology. (**b**) weight and atomic percentage of La, Sc, and Bi atoms. (**c**) The surface image shows the absence of grain boundaries. (**d**), (**e**), and (**f**) show EDX elemental color mapping for La, Sc, and Bi for the area in **c**, respectively.

## IV. Additional text

**Note 1:** The Kholer's rule of Hall resistivity.

Here we make a short introduction to the theory of the Kholer's rule. The Kholer's rule implies that

$$\frac{\rho - \rho_0}{\rho_0} = h(\frac{B}{\rho_0})$$

In semiclassical transportation theory, the conductivity tensor $\sigma_0$ is

$$\sigma_0 = \frac{ne^2\tau}{m^*}$$

while $n$ is the charge carrier density, $e$ is the electron charge, $\tau$ is the relaxation time and $m^*$ is the effective mass tensor. The resistivity $\rho_0 = \sigma_0^{-1}$ is

$$\rho_0 = \frac{m^*}{ne^2\tau}$$

we can get

$$MR \propto \rho\tau = f(B\tau)$$

$B\tau$ is combined variable, it determines the influence of magnetic field on magnetoresistance. It combines the influence of magnetic field and temperature together. All the $\rho - B$ curves at different temperatures can be scaled into a $\rho\tau - B\tau$ curve, which reflects the eigen properties of the magnetron response of the crystal[1].
For the Hall effects, the scaling behavior reads as

$$\frac{\rho_{yx}}{n\rho_0/m} = h(\frac{B}{n\rho_0/m})$$

it is called the Kohler's rule of Hall resistivity. Combined with the original Kohler's rule, it proposes a general way to deal with the carrier transportation in the magnetic field.

## V. Additional experimental and computational data

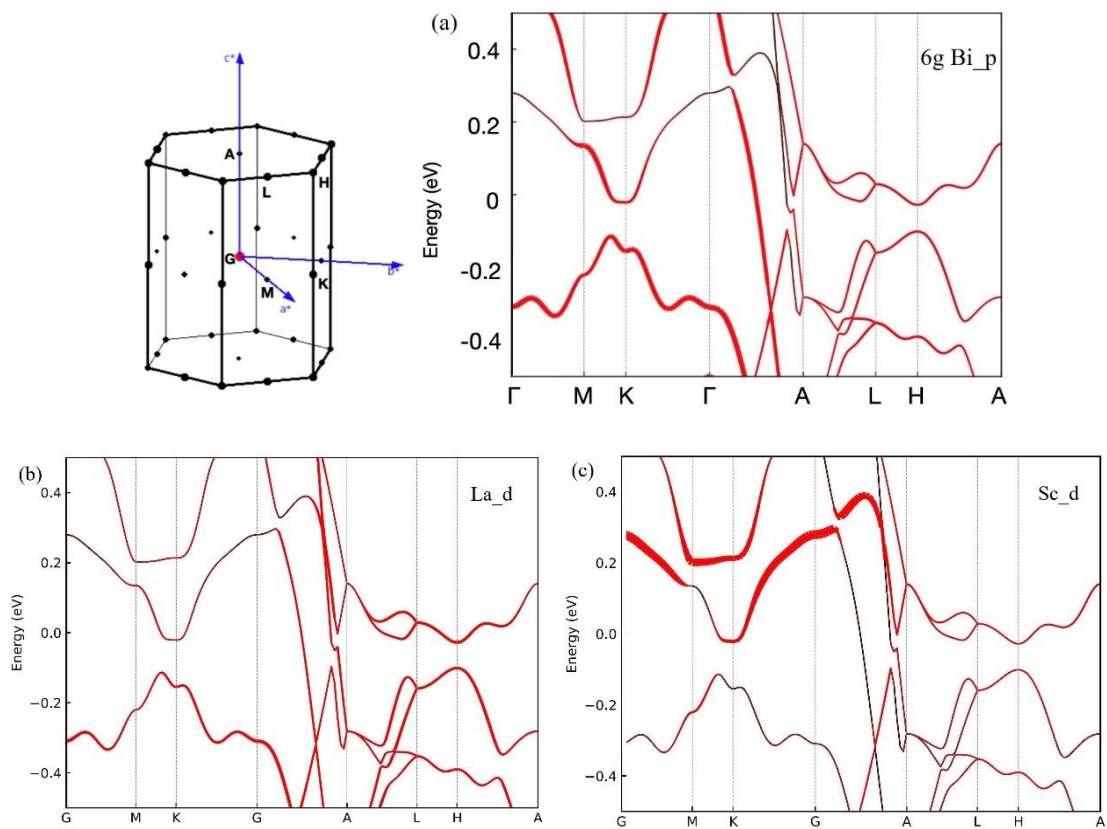

**Fig. S2** Projection band structure with the consideration of spin-orbit coupling at (**a**) *6g* Wyckoff positions of Bi, (**b**) La and (**c**) Sc. The color scale represents the contribution of the Bi *p* orbital in ScBi$_6$ face-sharing octahedral chains, the La *d* orbital and the Sc *d* orbital.

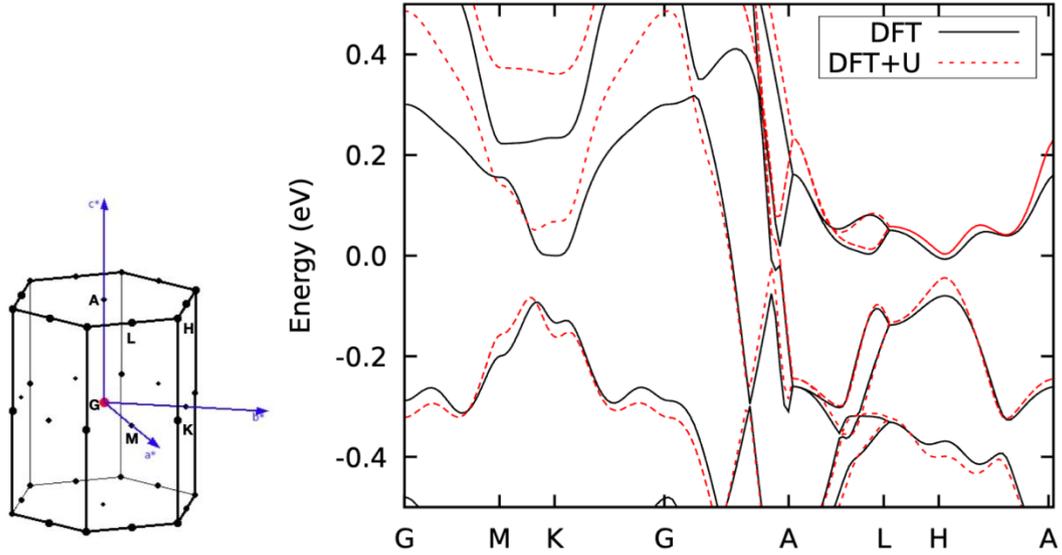

**Fig. S3 The electronic band structures of La$_3$ScBi$_5$ using the DFT method and the DFT+U method.** Implementing a Hubbard U correction of 2 eV on the d-orbitals of La atoms and 3 eV on the d-orbitals of Sc atoms in the DFT+U method[2-4] results in a band structure that exhibits modest deviation from that obtained via standard DFT methods, maintaining the same band trend. Notably, the band between the **Γ** and **A** points continues to display a steep slope, indicative of a high Fermi velocity.

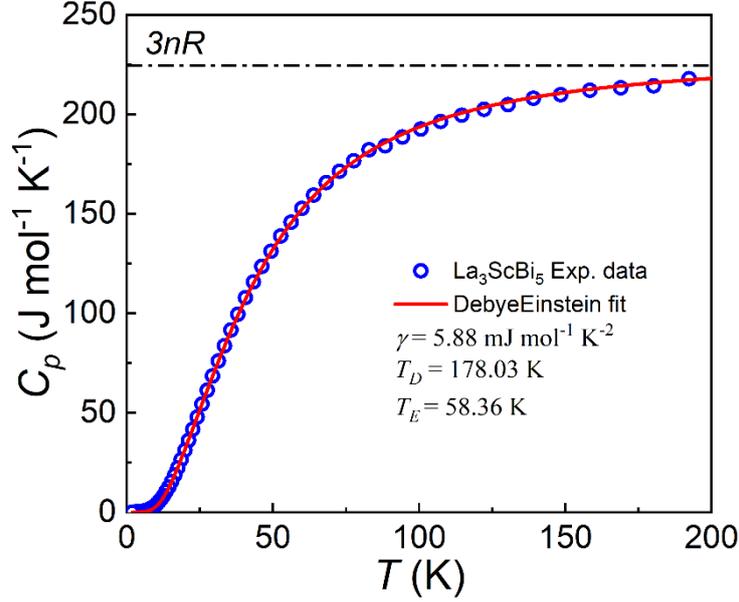

**Fig. S4** Temperature dependence of the heat capacity $C_p$ in the absence of a magnetic field.

To further investigate the specific heat properties of La$_3$ScBi$_5$, we employed the Debye-Einstein model to fit the data from 2 K to 300 K on the $C_p(T)$ curve, using the following analytical formula[5]:

$$C_p = \gamma T + 9nR\alpha \left(\frac{T}{T_D}\right)^3 \int_0^{\frac{T_D}{T}} \frac{x^4 e^x dx}{(e^x-1)^2} + 3nR(1-\alpha)\left(\frac{T_E}{T}\right)^2 \frac{e^{\frac{T_E}{T}}}{\left(e^{\frac{T_E}{T}}-1\right)^2}$$

In this formula, $\alpha$ serves as a weight factor between the Debye and Einstein models. The fitting results yield $T_D$ = 178.03 K, $T_E$ = 58.36 K, and $\gamma$ = 5.88 mJ mol$_{La}^{-1}$K$^{-2}$. The red curve depicted in the plot represents the combined fit of the lattice and electron contributions using the Debye-Einstein models. Notably, no phase transition is observed down to 2 K.

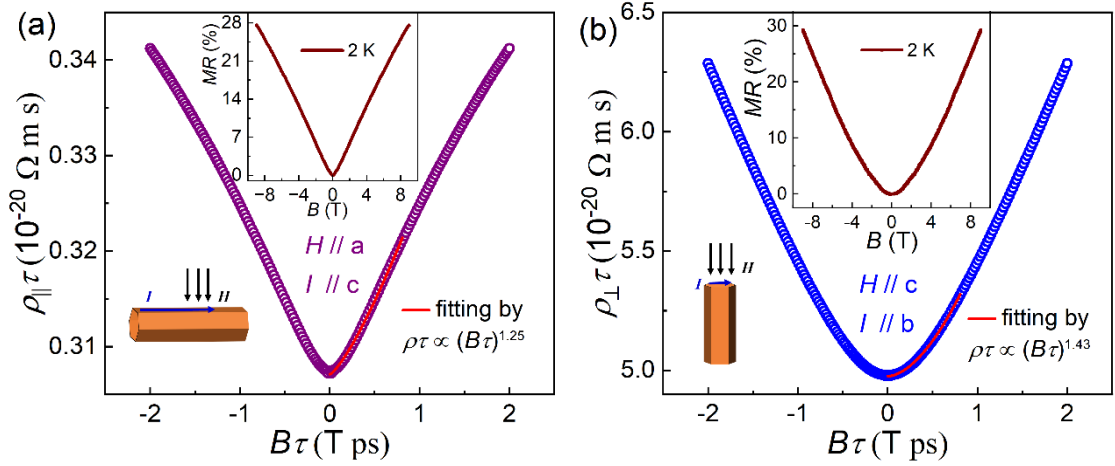

**Fig. S5 Calculated magnetoresistance.** (a) The calculated and measured MR at 2 K for *H // a* and *I // c* as a function of the magnetic field. Red lines represent the fitting data as a power function of the magnetic field. (b) The calculated and measured MR at 2 K for *H // c* and *I // b* as a function of the magnetic field. Red lines represent the fitting data as a power function of the magnetic field.

Calculation details of $\rho_\perp$ in Fig. S4b are as follows. On the one hand, the relaxation time is estimated using the equation: $\tau = (\rho_{\perp cal}\tau)_{B\tau=0}/\rho_{\perp exp} = 5 \times 10^{-20}/30 \times 10^{-8}\ s = 1/6 \times 10^{-12}\ s = 1/6\ ps$. On another hand, The calculated MR exhibits a linear dependence in the high-field region, reaching approximately 24% at a field strength of 9 T. Both the magnitude and trend of the calculated MR are in good agreement with the experimental observations. The calculated $\rho_\perp\tau$ shows a nonsaturating dependence on $(B\tau)^{1.43}$.

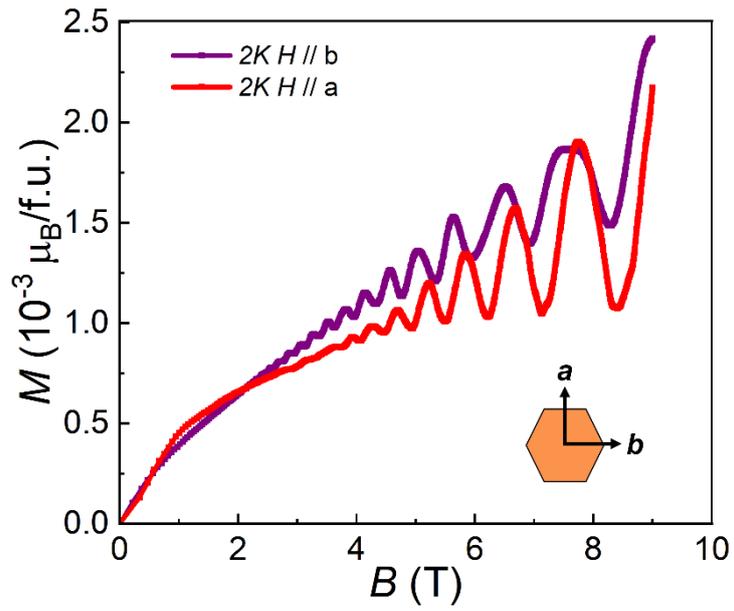

**Fig. S6 Isothermal magnetization of La$_3$ScBi$_5$ measured at 2 K in *ab*-plane.** The distinct and varying quantum oscillations between the *H* // a and the *H* // b orientation indicate significant in-plane anisotropy within the *ab*-plane of La$_3$ScBi$_5$.

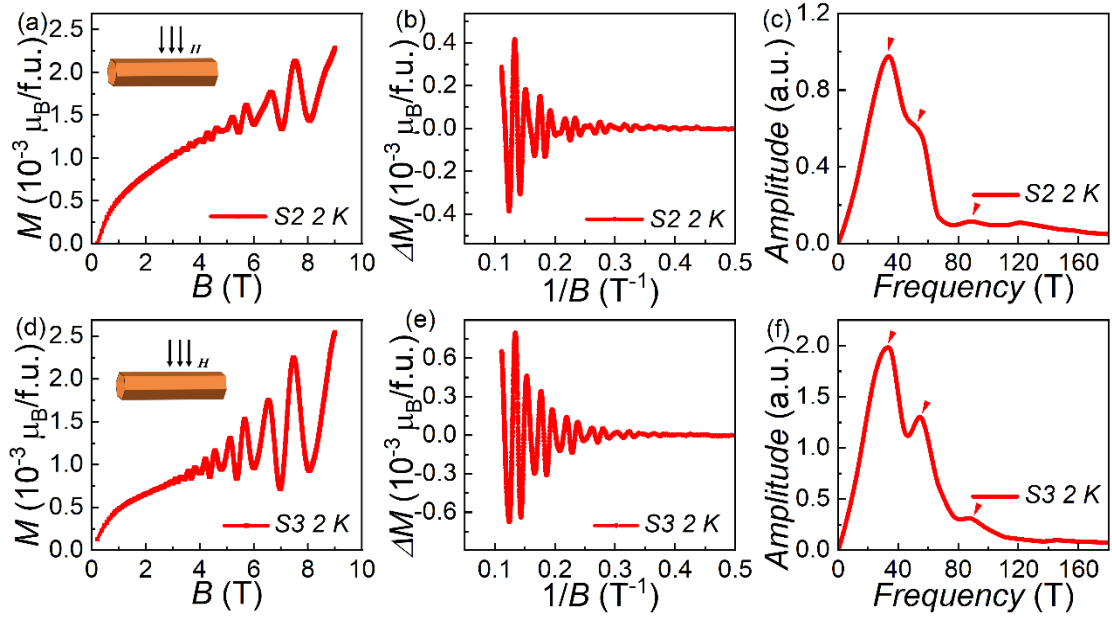

**Fig. S7** (**a**) and (**d**) Isothermal magnetization with $H // a$ at $T = 2$ K for sample S2, S3. (**b**) and (**e**) The oscillatory components of magnetization for S2 and S3. (**c**) and (**f**) The corresponding FFT spectrum of the oscillatory component of the dHvA oscillations at 2 K.

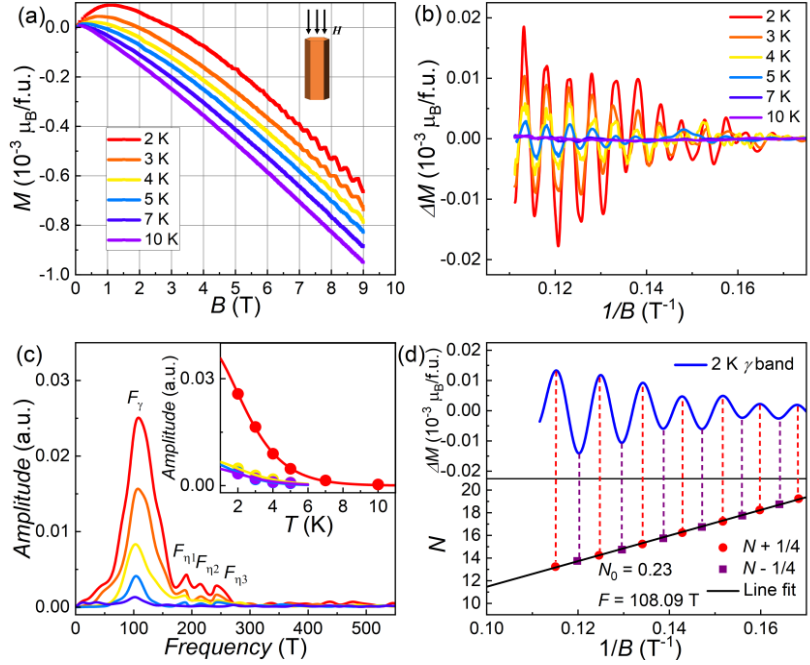

**Fig. S8 Analyses of the dHvA oscillations for *H // c* in La$_3$ScBi$_5$.** (**a**) Isothermal magnetization at different temperatures with the magnetic field parallel to the c-axis on sample S1. (**b**) The oscillatory components of magnetization *ΔM*. (**c**) The corresponding FFT spectrum of the oscillatory component of the dHvA oscillations at various temperatures. Inset: The fits of the FFT amplitudes to the temperature damping factor $R_T$ by the LK formula. (**d**) The LL index fan diagram for γ band derived from the oscillatory magnetization at 2 K.

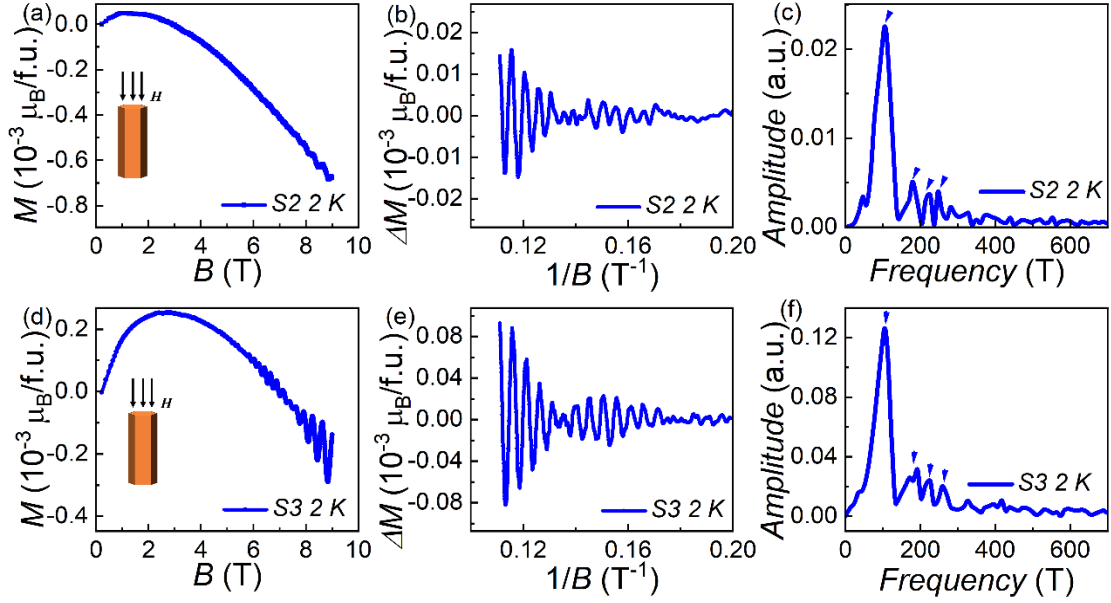

**Fig. S9** (**a**) and (**d**) Isothermal magnetization with $H \parallel c$ at $T = 2$ K for sample S2, S3. Note that the low-field regions of S2 and S3 show significant paramagnetic contributions near the zero field. (**b**) and (**e**) The oscillatory components of magnetization for S2 and S3. (**c**) and (**f**) The corresponding FFT spectrum of the oscillatory component of the dHvA oscillations at 2 K.

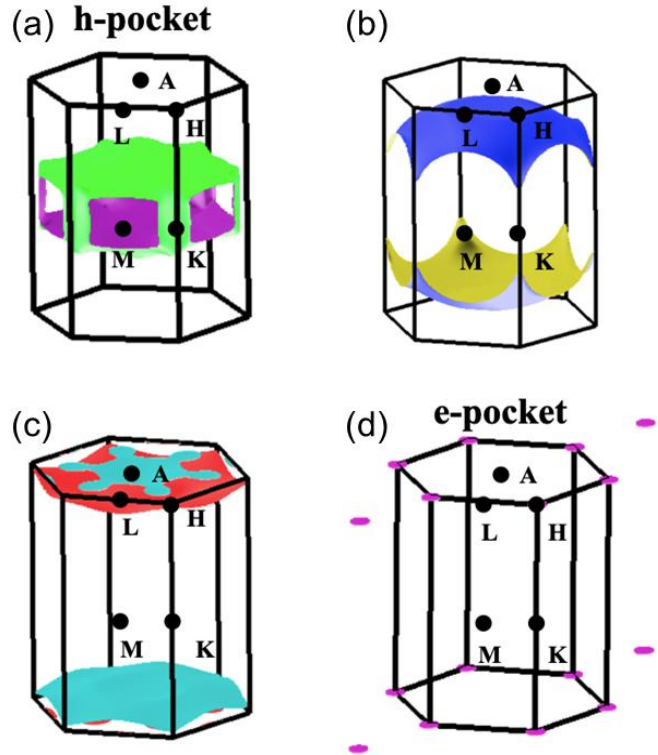

**Fig. S10 The Fermi surfaces in the presence of SOC.** Panels (**a**)-(**d**) illustrate the four pockets contributing to the dHvA oscillations when an external magnetic field is applied along the direction from the principal [100] to [001] axes. Here, *h* denotes the hole pockets, while *e* represents the electron pocket. The complexity of the Fermi surfaces leads to an angular dependence in the dHvA oscillation frequencies.

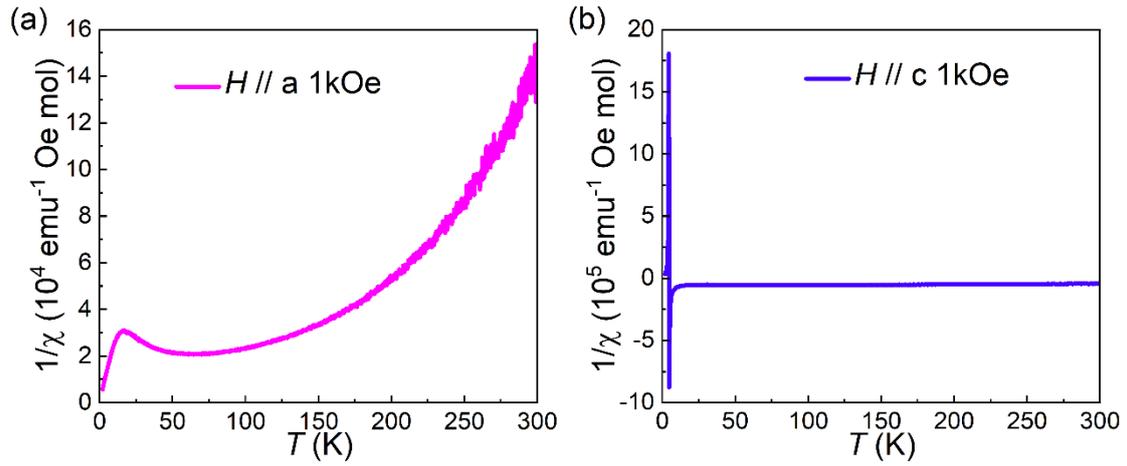

**Fig. S11.** The inverse susceptibility of La$_3$ScBi$_5$ under $H // a$ and $H // c$ ($H$ = 1 kOe) conditions, derived from the temperature-dependent susceptibility of sample S1, as shown in Fig. 7a. For the La$_3$ScBi$_5$ system with zero intrinsic magnetic moments of atoms, it appears challenging for the material to conform to the Curie law for both $\chi_a$ in the $H // a$ direction and $\chi_c$ in the $H // c$ direction.

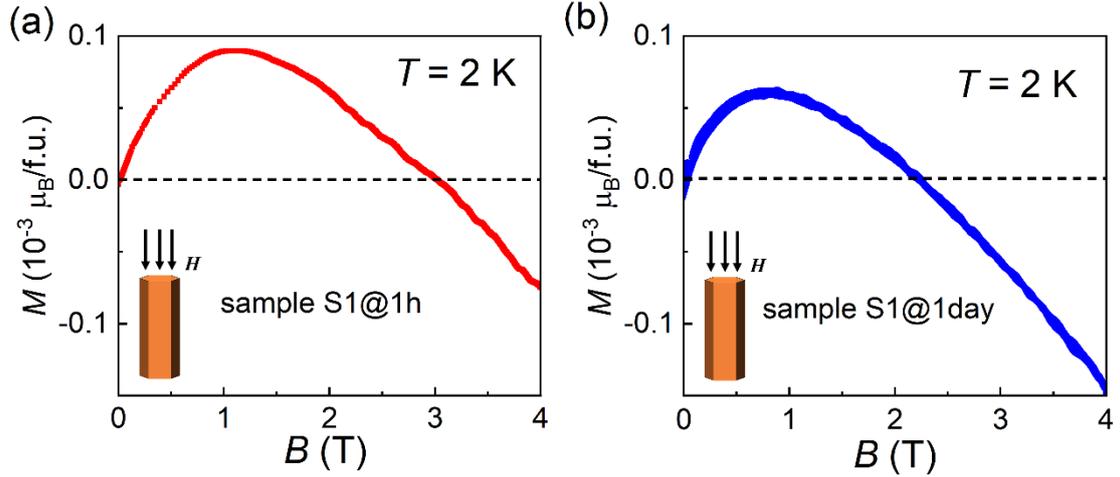

**Fig. S12.** Susceptibility cusp for a La$_3$ScBi$_5$ crystal on sample S1 at 2 K measured (**a**) an hour and (**b**) a day after the crystal growth. Fig. S12a shows data from Fig. S8, while Fig. S12b contains data from Fig. 7c. A clear aging effect was observed when testing the same sample multiple times, manifested as a decrease in the intensity of the paramagnetic anomaly over time. We tested the newly grown "fresh" sample S1 for dHvA oscillations associated with Fig. S8 for approximately 1 day, followed by isothermal $M(H)$ tests at different temperatures related to Fig. 7c. We found that the total cusp height in the low-field region decreased over time. This aging effect has also been reported in 3D topological materials[6] such as Bi$_2$Se$_3$, Bi$_2$Te$_3$, and Sb$_2$Te$_3$, reminiscent of aging effects observed in angle-resolved photoemission spectroscopy, where electronic structure near the surface undergoes "band bending," leading to the formation of a similar anisotropic state near the base of the bulk band[7, 8].

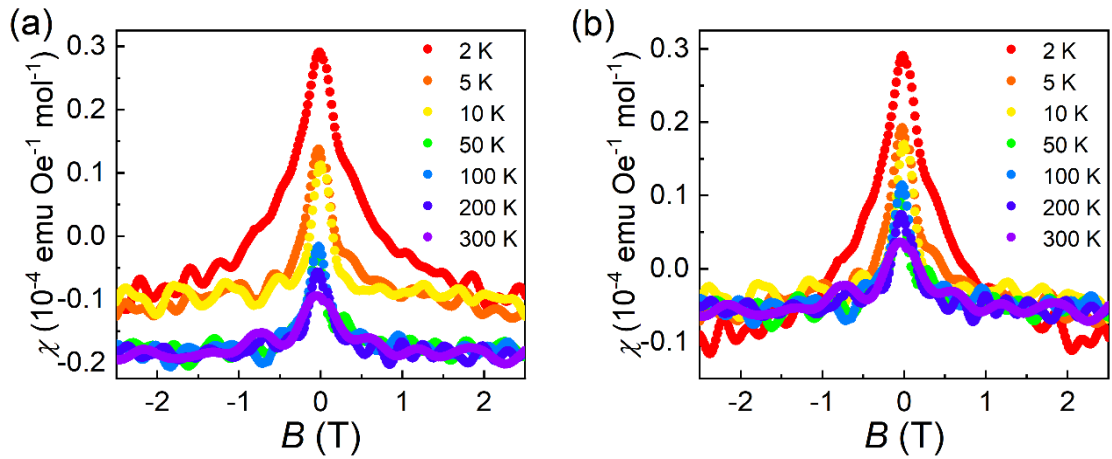

**Fig. S13.** The paramagnetic susceptibility cusp rides on a temperature dependent diamagnetic background, shown in (**a**) for $La_3ScBi_5$. It is considered that the diamagnetic response depends on the specifics of the band structure and the position of the chemical potential above $H = 0.5$ T[6]. In order to facilitate the analysis of peak characteristics, the high temperature data was shifted in Fig.7d of the manuscript, as shown in (**b**).

## Supplementary References: